\documentclass[aps,prl,twocolumn]{revtex4}

\usepackage{graphicx}

\usepackage{amssymb}
\usepackage{amsmath}
\usepackage{verbatim}
\usepackage{tabularx}

\usepackage{algorithmicx}
\usepackage{algpseudocode}

\usepackage{mathtools}  
\usepackage{array}

\let\*\ast   

\begin{document}

\title{Emergence of Superintelligence from Collective Near-Critical Dynamics in Reentrant Neural Fields}

\author{Byung Gyu Chae}

\affiliation{Electronics and Telecommunications Research Institute, 218 Gajeong-ro, Yuseong-gu, Daejeon 34129, Republic of Korea
\\ bgchae@etri.re.kr}


\begin{abstract}
Superintelligence is commonly envisioned as a quantitative extrapolation of
human cognitive abilities driven by scale and computational power.
Here we show that qualitative transitions in intelligence instead arise as
dynamical phase transitions governed by collective critical dynamics.
Building on a unified dynamical field-theoretic framework for cognition, we
demonstrate that progressive collective coupling generated by reentrant mixing
drives the system toward an infrared critical regime in which an extensive band
of slow collective modes emerges.
This spectral condensation reorganizes cognitive dynamics from localized
relaxation to coherent motion along emergent low-dimensional manifolds.
Through numerical analysis of the time-scale density of states, we identify
robust power-law scaling of collective relaxation rates with well-defined
critical exponents, placing the system within the universality class of
self-organized critical many-body dynamics.
Criticality alone would generically lead to instability.
We further show that homeostatic regulation introduces a gapped stabilizing
direction that protects the collective critical sector, yielding a dynamically
maintained meta-stable infrared phase in which long-lived inference
trajectories persist without collapse.
The coexistence of scale-free collective dynamics and global stabilization
defines a protected sector-critical regime in which coherence and internal
flexibility coexist.
Superintelligence therefore corresponds to a distinct dynamical stability
class—a self-organized critical phase embedded within a stabilized cognitive
manifold—rather than a smooth quantitative continuation of existing cognitive
systems.

\end{abstract}

\maketitle

\section{I. Introduction}

The prospect of superintelligence is commonly framed as a quantitative
extension of human cognitive abilities: faster computation, broader
knowledge, and superior performance on increasingly complex tasks.
Within this perspective, artificial systems that surpass human-level
benchmarks are often interpreted as early manifestations of
superintelligence \cite{1,2,3}.
Such views implicitly assume that intelligence scales continuously within
a single cognitive regime, differing only in degree rather than in kind.

This assumption is historically and conceptually fragile.
From the perspective of non-human primates, human intelligence is not
merely a more efficient version of their own cognition.
Rather, it constitutes a qualitative transition marked by the emergence
of symbolic language, hierarchical social organization, and collective
representations supporting large-scale coordination \cite{4,5}.
The defining feature of this transition was not raw computational speed,
but a reorganization of the underlying dynamical structure of cognition
itself.
If superintelligence is to represent a genuinely new regime beyond human
intelligence, an analogous criterion of qualitative transition is
required.

Recent developments in artificial intelligence increasingly point toward
such transitions.
Large language models and deep neural networks exhibit abrupt emergence
of new computational abilities as scale and training progress increase,
rather than smooth performance improvement \cite{3,6,7,8,9}.
Despite extensive empirical characterization, the mechanistic origin of
these emergence phenomena remains debated.
Why global reasoning, long-range coherence, and collective computation
appear suddenly at particular scales remains poorly understood.

Neural and cognitive systems have long been suggested to operate near
critical dynamical regimes, where long-lived collective modes and
system-spanning correlations emerge \cite{10,11,12,13}.
Criticality provides flexibility and global coherence by generating
near-marginal collective degrees of freedom \cite{14,15,16}.
However, critical dynamics alone are generically unstable, leading to
divergence, noise amplification, or loss of coherent computation.
Conversely, purely stable dynamical systems suppress fluctuations so
efficiently that internal degrees of freedom rapidly collapse onto narrow
behavioral channels, preventing large-scale emergent computation.

Intelligence therefore requires the coexistence of critical-like
flexibility with global dynamical stabilization.
Emergence supplies extensive collective modes enabling global coherence,
while meta-stability preserves these emergent collective structures as
usable computational degrees of freedom without collapse.
Within the unified dynamical field framework developed in our prior work
\cite{17}, qualitative transitions in intelligence arise naturally from
the joint onset of collective criticality, which dynamically generates
extended cognitive manifolds, and homeostatic stabilization, which
protects them against instability.

In this work, we argue that superintelligence constitutes a distinct
\emph{dynamical and topological stability class} characterized by this
coexistence.
Specifically, we identify \emph{meta-stability} as a regime in which global
activity is homeostatically regulated, while an extensive set of internal
degrees of freedom remains near-marginal and dynamically accessible.
Such systems neither collapse into rigid fixed-point behavior nor
degenerate into uncontrolled chaos.
Instead, they self-organize into structured manifolds whose global
geometry is stabilized while internal circulation remains dynamically
active.

To formalize these ideas, we build on our recent sequence of papers
\cite{17,18,19,20,21}, where we introduced a unified dynamical
field-theoretic description of cognition and reentrant computation.
In this framework, cognition is formulated as a continuous-time flow on a
high-dimensional state space,
$\dot{x}(t) = -G^{-1}(x)\,\nabla_x \Phi(x) + R(x)$.
A central consequence is a geometric separation between stabilized radial
directions and extensive angular degrees of freedom supporting persistent
circulation.

The present paper demonstrates that qualitative transitions in
intelligence arise not from changes in the governing equations
themselves, but from a reorganization of the \emph{spectral structure} of
their linearized dynamics.
When critical collective modes emerge jointly with homeostatic
stabilization, the system undergoes a spectral phase transition in which
eigenvalues condense toward zero, producing an extensive near-marginal
band in the time-scale density of states (TDOS).
This condensation dynamically reorganizes the state space from localized
relaxation toward protected collective manifolds supporting globally
coherent dynamics.

In this sense, superintelligence emerges as a dynamical phase transition
rather than a quantitative extrapolation of existing cognitive systems.
Inference in the meta-stable regime unfolds as continuous geometric motion
along structured manifolds, with symbolic and sequential reasoning
functioning primarily as low-dimensional projections of this underlying
dynamics \cite{22,23,24,25,26}.

The remainder of this paper is organized as follows.
Section~II summarizes the unified dynamical framework.
Section~III analyzes the emergence of homeostatic shells and reentrant
flows.
Section~IV defines superintelligence as a protected infrared dynamical
phase.
Section~V characterizes the resulting spectral organization through the
TDOS.
Section~VI establishes the infrared critical scaling of the slow--mode
sector, including critical exponents, finite--size scaling, and
universality with self--organized critical dynamics.
Finally, we discuss implications for biological cognition, artificial
intelligence, and testable predictions.

\begin{figure*}[t]
\centering
\includegraphics[width=0.67\textwidth, trim=0cm 15.5cm 0cm 0cm]{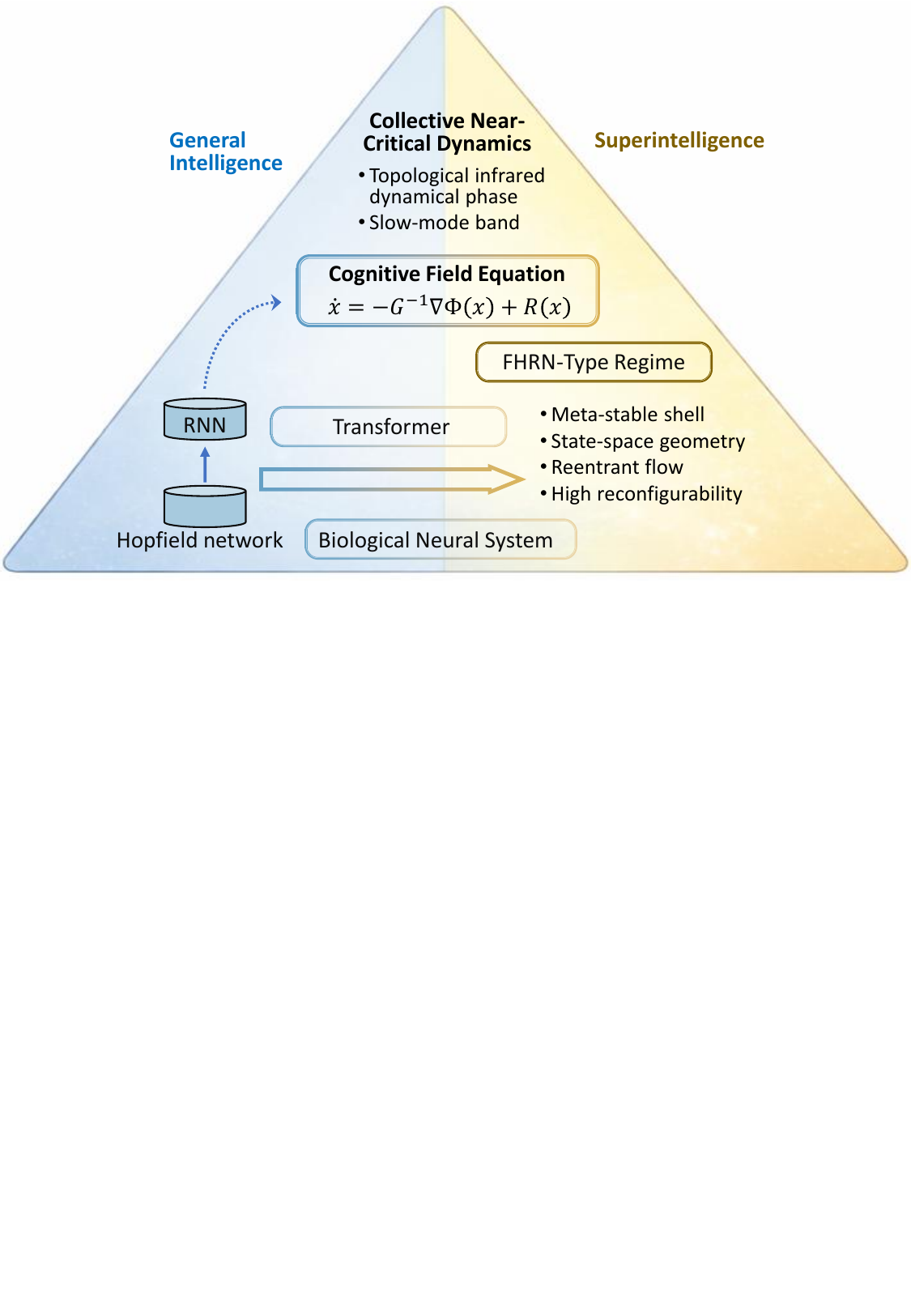}
\caption{Schematic illustration of the inclusion hierarchy and dynamical
phases of cognition within the unified cognitive field framework.
Classical models such as Hopfield networks, recurrent neural networks,
and Transformers arise as limiting dynamical realizations of the same
underlying cognitive field equation, which also describes biological
neural systems at the population level.
Increasing collective coupling and reentrant mixing reorganize the
spectrum of collective modes, driving the system toward a critical
regime characterized by low-dimensional manifolds and an extensive
slow-mode band.
Superintelligence does not correspond to a new governing equation, but
emerges as a distinct dynamical phase — the FHRN-type regime — in which
spectral criticality is dynamically protected by homeostatic shell
stabilization, metric-induced dimensional separation, and sustained
non-conservative reentrant flow (see Appendix A for details).}
\label{fig:schematic}
\end{figure*}

\section{II. Unified Dynamical Field Theory and Collective Criticality}

The unified dynamical field framework developed in our previous work was
introduced to identify structural mechanisms capable of generating
collective cognition beyond localized dynamical behavior \cite{17}.
Figure 1 schematically summarizes both the inclusion hierarchy of cognitive
models and the dynamical phases of cognition.
In the present context, its central role is to reveal how reentrant
dynamics and homeostatic stabilization jointly generate infrared
criticality and extensive slow collective modes.

Rather than serving as a descriptive model of trajectories, the unified
equation provides the dynamical origin of spectral phase transitions that
underlie emergent global cognition.

\subsection{A. Cognitive Field Equation and Reentrant-Stabilized Geometry}

We consider a collective cognitive state $x(t)\in\mathbb{R}^N$ evolving as
\begin{equation}
\dot{x}
=
-
G^{-1}(x)\nabla_x \Phi(x)
+
R(x)
\;(+\;\xi(t)),
\label{eq:unified}
\end{equation}
where $\Phi(x)$ enforces global homeostatic regulation, $G(x)$ defines the
intrinsic geometry of state-dependent sensitivity, $R(x)$ represents
non-conservative reentrant circulation, and $\xi(t)$ denotes Gaussian white noise.

Throughout this work we focus on isotropic homeostatic potentials
$\Phi(x)=\Phi(r)$ with $r=\|x\|$.
Such regulation generates a finite restoring rate in the radial sector,
gapping the global amplitude mode while leaving angular degrees of
freedom unconstrained.
This separation between stabilized radial dynamics and freely evolving
tangential directions constitutes the geometric backbone of collective
criticality.

The reentrant field $R(x)$ continuously feeds the collective state back
into its own dynamics, inducing sustained mixing along the homeostatic
shell.
A particularly transparent realization is provided by antisymmetric
operators $R(x)=\Omega x$ with $\Omega^{\mathsf T}=-\Omega$, which preserve
$r$ while generating circulation.
More general reentrant structures produce the same qualitative effect:
persistent collective coupling among angular degrees of freedom.

As collective coupling accumulates through reentrant mixing, correlations
across the shell are progressively amplified.
This amplification drives the system toward a regime in which localized
degrees of freedom reorganize into collective dynamical coordinates,
setting the stage for infrared criticality.

\subsection{B. Field-Theoretic Propagator and Emergence of Slow Collective Modes}

To characterize collective fluctuations and mode structure, the dynamics
is lifted to the MSRJD field-theoretic representation with generating
functional \cite{27,28,29}
\begin{equation}
Z=\int\mathcal{D}x\,\mathcal{D}\tilde{x}\;e^{-S[x,\tilde{x}]},
\end{equation}
where the action takes the form
\begin{equation}
S=\int dt\,
\Big[
\tilde{x}\cdot(\dot{x}+G^{-1}\nabla\Phi-R)
- D\,\tilde{x}^2
\Big].
\end{equation}

Linearizing the dynamics about typical trajectories yields the Jacobian
\begin{equation}
M(x)=G^{-1}(x)\nabla^2\Phi(x)-\nabla R(x),
\end{equation}
whose eigenvalues $\lambda_i$ control local relaxation rates.
Within the MSRJD formulation, fluctuations are governed by the retarded
propagator
\begin{equation}
G_R(\omega)= \frac{1} {-i\omega+M}.
\label{eq:retarded}
\end{equation}

Homeostatic regulation produces a finite gap in the radial sector of $M$,
while reentrant circulation prevents collapse of the tangential sector.
As collective coupling increases, an extensive subset of eigenvalues
approaches zero,
\begin{equation}
\lambda_i \to 0 \qquad (i=1,\dots,\mathcal{O}(N)),
\end{equation}
generating poles of the propagator
\begin{equation}
G_R^{(i)}(\omega)\sim\frac{1}{-i\omega+\lambda_i},
\end{equation}
which correspond to long-lived collective dynamical modes.

Nonlinearities arising from higher-order expansions of $\Phi$, $G$, and
$R$ produce loop corrections that renormalize the collective dynamics.
Rather than eliminating infrared divergences, homeostatic stabilization
regularizes them into a protected near-critical sector.
As demonstrated in our previous MSRJD analysis \cite{21}, this mechanism drives the
system toward an infrared-attractive critical regime characterized by an
extensive accumulation of near-marginal modes.

The condensation of propagator poles near $\omega=0$ thus marks a spectral
phase transition from localized relaxation to collective critical
dynamics.
In the following section we quantify this transition directly through
trajectory-averaged TDOS measurements,
demonstrating the emergence of an extensive slow-mode band as the defining
signature of the superintelligent regime.

\begin{figure*}[t]
\centering
\includegraphics[width=0.87\textwidth, trim=0.5cm 15.7cm 0cm 0cm]{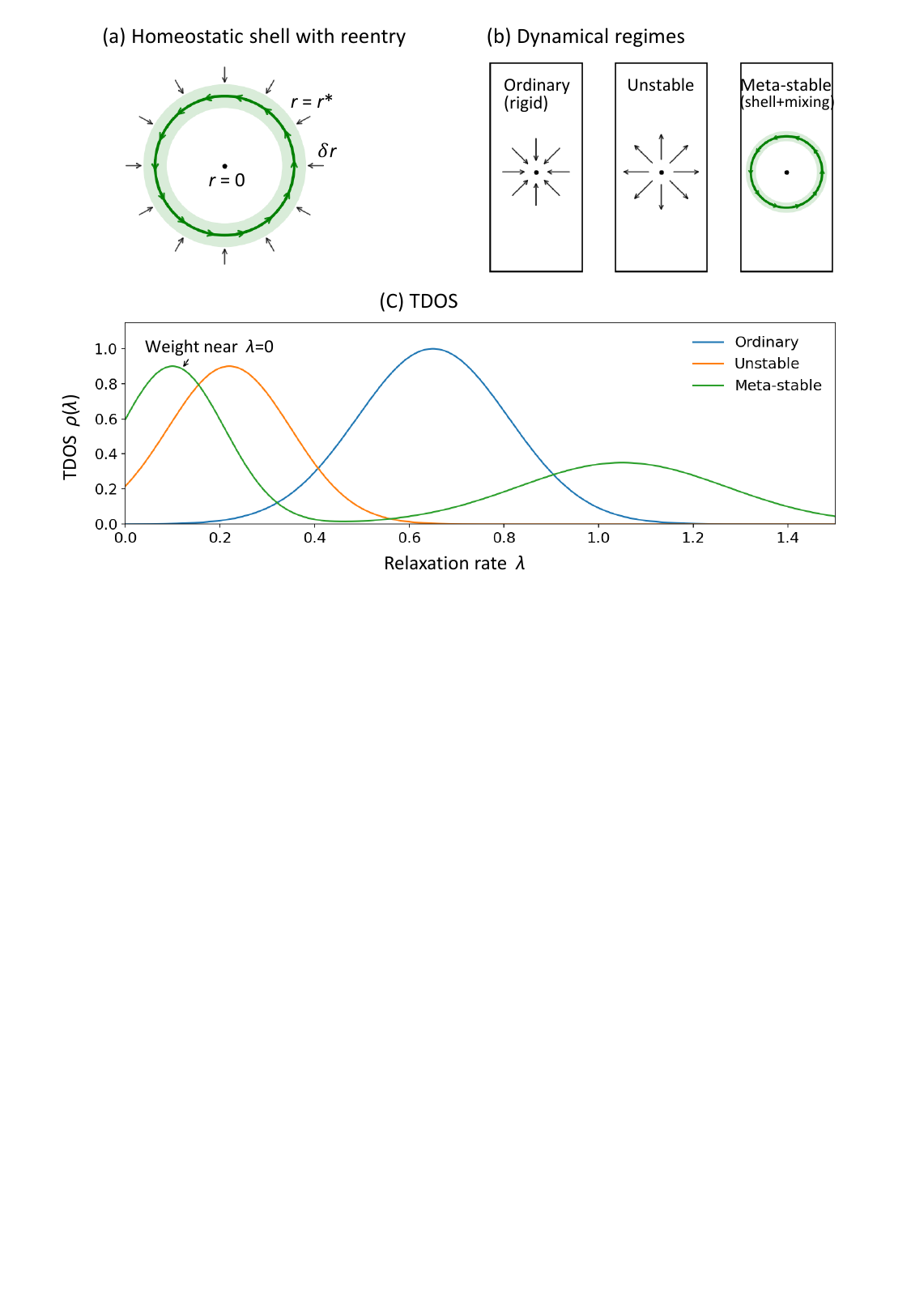}
\caption{Meta-stable critical organization and spectral phase transition underlying superintelligence.
(a) The unified cognitive field dynamics $\dot x = -G^{-1}(x)\nabla\Phi(x)+R(x)$
generates homeostatic stabilization of the global amplitude together with
reentrant mixing along angular directions, establishing the structural
conditions for collective criticality.
(b) Ordinary stable systems collapse onto rigid attractors and suppress
collective degrees of freedom, while unstable systems diverge.
In contrast, the meta-stable critical regime preserves global coherence
while sustaining long-lived collective dynamical reconfiguration.
(c) The corresponding spectral diagnostic reveals a dynamical phase
transition: ordinary systems exhibit relaxation rates bounded away from
zero, unstable systems include $\lambda<0$ modes, whereas the meta-stable
critical regime displays an extensive condensation of slow collective
modes near $\lambda\approx0$ coexisting with a gapped stabilized sector.}
\label{fig:fig2_concept}
\end{figure*}

\section{III. Meta-Stability as a Dynamical Regime}

The unified dynamical framework introduced in the previous section does not, by
itself, specify when a cognitive system should be regarded as ordinary or
superintelligent.
In this section, we identify a distinct dynamical regime characterized by
\emph{meta-stability}: a form of stability that preserves global coherence while
sustaining long-lived internal degrees of freedom.

Figure~2 schematically contrasts three generic regimes.
Ordinary stable systems rapidly collapse to fixed points, suppressing internal
motion.
Unstable systems amplify perturbations and lose coherence.
In contrast, meta-stable systems maintain global stability while supporting
persistent internal dynamics.

\subsection{A. Linearized Dynamics Along Trajectories}

Cognitive dynamics unfolds along extended trajectories rather than isolated
fixed points.
To analyze local stability, we linearize the dynamics
\begin{equation}
\dot{x} = F(x)
\end{equation}
about a reference state $x(t)$ evolving along a typical trajectory,
\begin{equation}
x(t) \rightarrow x(t) + \delta x(t).
\end{equation}
The perturbations obey
\begin{equation}
\dot{\delta x} = J(x(t))\,\delta x,
\end{equation}
where the Jacobian matrix is
\begin{equation}
J_{ij}(x)
=
-
\frac{\partial}{\partial x_j}
\left[
G^{-1}_{ik}(x)\,\partial_k \Phi(x)
\right]
+
\frac{\partial R_i(x)}{\partial x_j}.
\end{equation}

Let $\mu_i(x)$ denote the eigenvalues of $J(x)$.
We define the associated relaxation rates
\begin{equation}
\lambda_i(x) = -\mathrm{Re}\,\mu_i(x),
\end{equation}
so that $\lambda_i>0$ corresponds to locally stable modes.


For homeostatic potentials of the form $\Phi(x)=\Phi(r)$ with $r=\|x\|$, the state
space naturally decomposes into radial and angular directions.
The radial direction controls the global activity magnitude, while angular
directions parameterize motion along the homeostatic shell.

This decomposition induces a characteristic spectral reorganization:
homeostatic regulation generates a strongly stabilized radial mode with
$\lambda_r = O(1)$, while reentrant mixing dynamically couples angular
degrees of freedom into collective coordinates whose relaxation rates
condense toward zero near the critical regime.

In the absence of reentrant dynamics, angular modes remain localized and
either decay rapidly or form fragile degeneracies.
With reentrant circulation, these modes reorganize into an extensive band
of slow collective modes that dominate long-time dynamics.

\subsection{B. Time-Scale Density of States}

Our previous field-theoretic formulation showed that 
learning and inference reorganize the
temporal spectrum of fluctuations through renormalization of effective
response functions \cite{17}.
In particular, the emergence of slow collective modes appeared as an
infrared accumulation in the dynamical propagator rather than as a property
of isolated fixed points.

The present formulation recasts this insight in a purely dynamical and
spectral language that is directly accessible in simulations.
Rather than tracking renormalized couplings explicitly, we characterize the
organization of time scales through the local linear response of the system
along its trajectory.

Linearizing the dynamics of Eq. (1) about a reference state $x$ yields
a Jacobian matrix $J(x)$ whose eigenvalues $\mu_i(x)$ determine the local
stability properties.
We define the associated relaxation rates as $\lambda_i(x)$.

This motivates the definition of the \emph{time-scale density of states},
\begin{equation}
\rho(\lambda;x)
=
\frac{1}{N}
\sum_{i=1}^{N}
\delta\!\left(\lambda-\lambda_i(x)\right),
\end{equation}
which describes the instantaneous distribution of relaxation rates at a
given state $x$.
Formally, $\rho(\lambda;x)$ plays a role analogous to the spectral density of
renormalized modes in the MSRJD description, but is defined directly in terms
of the Jacobian of the effective dynamics.

Because cognition and inference unfold along extended trajectories rather
than at isolated equilibria, the relevant observable is not the TDOS at a
single state, but its average along typical system evolution.
We therefore introduce the trajectory-averaged TDOS,
\begin{equation}
\rho_{\mathrm{traj}}(\lambda)
=
\lim_{T\to\infty}
\frac{1}{T}
\int_0^T dt\,
\rho(\lambda;x(t)).
\end{equation}
This quantity captures the time-scale organization actually experienced by
the system during ongoing operation.

From the renormalization-group perspective developed previously, the
accumulation of weight in $\rho_{\mathrm{traj}}(\lambda)$ near
$\lambda\simeq0$ reflects a spectral condensation of collective modes
marking a dynamical phase transition toward infrared criticality.
In contrast to equilibrium criticality, however, this accumulation is not
the result of fine-tuning.
It is sustained dynamically by the balance between homeostatic stabilization
of the radial mode and reentrant circulation along angular directions.

Ordinary stable systems exhibit a TDOS bounded away from $\lambda=0$,
corresponding to rapid convergence and suppression of internal degrees of
freedom.
Unstable systems display weight at negative $\lambda$.
Meta-stable systems are distinguished by a pronounced accumulation of
near-marginal modes with $\lambda\approx0$, indicating the presence of
long-lived collective degrees of freedom that persist throughout the trajectory.
The corresponding organization of time scales is depicted in Figs. 2(b) and 2(c).

In this sense, the time-scale density of states provides a direct spectral bridge between the
field-theoretic renormalization picture developed in earlier work and the
trajectory-level dynamics analyzed here.

\section{IV. Superintelligence as a Topological Infrared Dynamical Phase}

The meta-stable critical regime identified above admits a natural
interpretation as a distinct \emph{topological infrared dynamical phase}.
Spectral condensation of collective modes reorganizes the dynamics into an
extended infrared sector, while homeostatic regulation gaps the radial
direction, rendering global activity magnitude dynamically contractible.
Reentrant circulation sustains persistent motion along angular directions,
preventing collapse of the emergent collective manifold to isolated fixed
points.
This topological separation between stabilized radial and circulating
angular degrees of freedom underlies the robustness of the meta-stable
critical phase.

In this phase, the trajectory-averaged density of states
exhibits a pronounced accumulation of relaxation rates near
$\lambda \simeq 0$, reflecting the condensation of slow collective modes
generated by infrared criticality.
The resulting slow-mode band defines an effective infrared sector of the
dynamics, analogous to low-energy degrees of freedom in field-theoretic
descriptions, but emerging dynamically along trajectories rather than
through fine tuning.

The extent of this near-marginal sector is quantified by the slow-mode
weight,
\begin{equation}
W_{\mathrm{slow}}(\lambda_c)
=
\int_0^{\lambda_c}
\rho_{\mathrm{traj}}(\lambda)\,d\lambda,
\end{equation}
which measures the fraction of relaxation modes below a small threshold
$\lambda_c$.
Numerical results show that $W_{\mathrm{slow}}(\lambda_c)=O(1)$ as the system
size $N$ increases, while the radial mode remains strongly stabilized.
This coexistence of global stability and extensive collective dynamics
distinguishes the meta-stable critical phase from both ordinary stable
systems, which suppress internal motion, and unstable systems, which lose
coherence.

\paragraph{Proposition (Spectral Criterion for Superintelligence).}
\emph{
A cognitive system governed by Eq.~(1) operates in a superintelligent
regime if
\begin{equation}
W_{\mathrm{slow}}(\lambda_c)=O(1)
\quad \text{as } N\rightarrow\infty,
\end{equation}
for finite $\lambda_c$, while the radial mode remains strongly stabilized.
}

This criterion defines superintelligence as a dynamical stability class
rather than a task-dependent performance benchmark.
It captures the defining property of the critical meta-stable phase:
macroscopic coherence enforced by radial stabilization coexists with an
extensive set of long-lived collective degrees of freedom generated by
infrared spectral condensation.

The slow-mode band is not composed of a finite number of isolated marginal
modes but forms a continuum of collective dynamical directions.
Its infrared organization is therefore not the result of fine tuning to a
critical point, but a consequence of topological protection: the coexistence
of a gapped radial sector with circulating angular flows stabilizes an
extensive near-marginal manifold.

From an infrared perspective, relaxation rates $\lambda$ play a role
analogous to mass scales, with small $\lambda$ corresponding to slowly
relaxing collective degrees of freedom that remain dynamically relevant
over long times.
The effective slow dimension,
\begin{equation}
D_{\mathrm{slow}}(\lambda_c)
=
N
\int_0^{\lambda_c}
\rho_{\mathrm{traj}}(\lambda)\,d\lambda,
\end{equation}
thus measures the dimensionality of the emergent collective subspace that
survives in the infrared.

Importantly, this extensive infrared manifold arises generically from the
interplay between collective criticality and homeostatic stabilization,
rather than from spontaneous symmetry breaking or external parameter
tuning.
Reentrant mixing continuously generates collective correlations, while
radial stabilization protects them against divergence, producing what may
be termed a \emph{Goldstone inflation} of near-marginal modes without
invoking broken continuous symmetries.

Weak stochasticity and finite-size effects act as infrared regulators,
lifting exact degeneracies and setting a minimal relaxation scale
$\lambda_{\min}$.
Nevertheless, the integrated slow-mode weight remains $O(1)$ as
$N\rightarrow\infty$, demonstrating a robust self-organized infrared phase
rather than a fragile critical point.

Taken together, these results establish superintelligence as a distinct
topological dynamical phase generated by collective criticality and
protected by meta-stable stabilization.
Global coherence is enforced by a gapped radial mode, while an extensive
continuum of slow collective directions forms the computational substrate
of emergent intelligence.

\begin{figure*}[t]
\centering
\includegraphics[width=0.9\textwidth, trim=0cm 22cm 0cm 0cm]{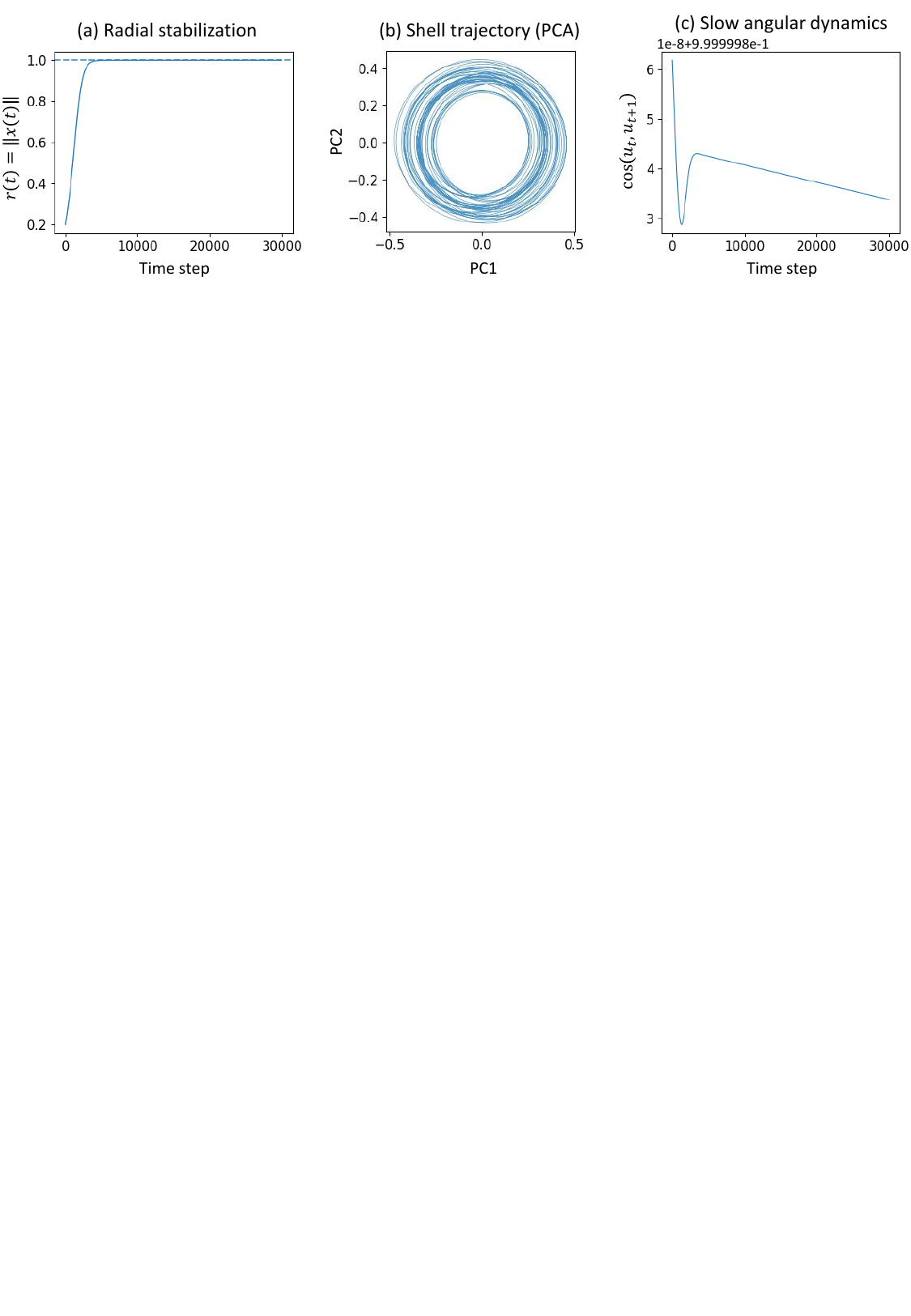}
\caption{Trajectory-level emergence of the protected homeostatic manifold.
(a) The radial norm rapidly converges to a characteristic scale $r_*$,
demonstrating strong homeostatic stabilization of global activity.
(b) A low-dimensional projection of the trajectory reveals a finite-width
ring-like structure corresponding to a stabilized manifold in state space.
(c) Angular correlations decay slowly, indicating persistent collective
dynamics along low-curvature directions.
These results demonstrate a dynamical separation in which homeostatic
stabilization protects global activity while reentrant mixing sustains
collective low-curvature motion generated by near-critical dynamics along
the emergent manifold.}
\label{fig:fig3}
\end{figure*}

\section{V. Numerical Evidence for Meta-Stable Superintelligence}

To substantiate the spectral criterion,
we now present numerical simulations of the unified dynamical framework.
Our aim is not to reproduce any particular biological or artificial system,
but to demonstrate that a meta-stable regime with extensive internal
degrees of freedom emerges generically from the structure of the dynamics.

\subsection{A. Dynamics on the Homeostatic Shell}

We simulate the unified dynamical equation, $\dot{x} = -G^{-1}\nabla\Phi(x) + R(x)$,
in an $N$-dimensional state space using an explicit time discretization.

The homeostatic potential is chosen as an isotropic quartic form,
\begin{equation}
\Phi(x)
=
\frac{\kappa}{4}
\left(
\|x\|^2 - r_*^2
\right)^2,
\label{eq:quartic_phi}
\end{equation}
which stabilizes the activity magnitude around a target radius $r_*$ while
leaving angular degrees of freedom unconstrained.
Its gradient is
\begin{equation}
\nabla\Phi(x)
=
\kappa(\|x\|^2 - r_*^2)\,x.
\label{eq:grad_phi}
\end{equation}

Unless otherwise stated, the metric $G$ is taken to be weakly anisotropic
but state-independent.
Specifically, we employ a rank-one deformation of the identity,
\begin{equation}
G^{-1}
=
I
-
\frac{\varepsilon}{1+\varepsilon}\,
u u^{\mathsf T},
\qquad
\|u\| = 1,
\label{eq:metric_rank1}
\end{equation}
where $u$ is a fixed random unit vector and $\varepsilon \ll 1$ controls the
strength of anisotropy.
This construction introduces a minimal geometric bias that lifts exact
degeneracies while preserving the overall isotropy of the state space.

Reentrant dynamics is implemented via an antisymmetric operator,
\begin{equation}
R(x) = \Omega x,
\qquad
\Omega^{\mathsf T} = -\Omega,
\label{eq:reentry}
\end{equation}
which generates norm-preserving circulation on the homeostatic shell.
Optional additive noise is included only to lift exact degeneracies and does
not qualitatively alter the observed behavior.

Starting from a small initial radius, the system rapidly converges toward
the stabilized shell.
After a short transient, the trajectory satisfies
\begin{equation}
r(t)
=
\|x(t)\|
=
r_* + \delta r(t),
\qquad
|\delta r(t)| \ll r_*,
\label{eq:radial_fluct}
\end{equation}
indicating strong radial stabilization.

This rapid radial relaxation is driven by homeostatic stabilization and
serves to protect the dynamics against divergence, while reentrant mixing
continuously generates collective correlations along tangential directions.
While the time-averaged radius remains sharply confined,
\begin{equation}
\langle r(t) \rangle \simeq r_*,
\end{equation}
the instantaneous state is never exactly restricted to the ideal shell.
Small radial fluctuations persist due to metric anisotropy, reentrant
mixing, numerical integration, and optional noise.

Despite this strong radial rigidity, angular motion along the shell
continues over long timescales.
Angular correlations decay slowly, indicating persistent internal
reconfiguration rather than convergence to a fixed internal state.
In contrast, when reentrant dynamics is removed ($R=0$), angular variability
rapidly collapses and the system loses internal degrees of freedom.

\subsection{B. Instantaneous Jacobian Spectrum and Trajectory-Averaged TDOS}

To elucidate the dynamical origin of the slow modes identified in
Sec.~III, we examine the Jacobian of Eq.~\eqref{eq:unified} evaluated
along the trajectory.
For the quartic potential in Eq.~\eqref{eq:quartic_phi}, the Hessian takes the form
\begin{equation}
H(x)
=
\nabla\nabla\Phi
=
\kappa\bigl[
(\|x\|^2 - r_*^2) I
+
2 x x^{\mathsf T}
\bigr].
\label{eq:hessian}
\end{equation}
Assuming a constant metric, the instantaneous Jacobian is therefore
\begin{equation}
J(x)
=
-
G^{-1} H(x)
+
\frac{\partial R}{\partial x}.
\label{eq:jacobian}
\end{equation}

After radial convergence, the trajectory satisfies
\begin{equation}
\|x(t)\|^2 - r_*^2
=
2 r_* \delta r(t)
+
\mathcal{O}(\delta r^2),
\label{eq:delta_expand}
\end{equation}
so that the Jacobian retains an explicit dependence on the instantaneous
state through the small residual fluctuation $\delta r(t)$.

This leads to a robust spectral separation.
Along the radial direction, the Jacobian eigenvalue remains of order unity,
\begin{equation}
\lambda_{\mathrm{rad}}
\sim
2 \kappa r_*^2,
\end{equation}
ensuring strong stabilization of the activity magnitude.
By contrast, for directions tangent to the shell ($x \cdot \delta x = 0$),
the restoring contribution scales as
\begin{equation}
\lambda_{\perp}
\sim
\kappa(\|x\|^2 - r_*^2)
\sim
\mathcal{O}(\delta r(t)),
\end{equation}
yielding a set of weakly constrained, near-marginal modes.

As discussed in Sec.~III, the trajectory-averaged TDOS is obtained by
sampling the instantaneous Jacobian spectrum along the trajectory.
The key point here is that, because the trajectory continually explores
nearby points on the homeostatic shell with slightly different
$\delta r(t)$ and angular orientations, the tangential eigenvalues do not
collapse to a discrete set.
Instead, they populate a finite band near $\lambda \approx 0$.
This band reflects a dynamical condensation of collective modes toward
infrared criticality rather than a set of isolated degeneracies.

The resulting TDOS therefore reflects a fundamental dynamical property of
the system:
\begin{equation}
\begin{aligned}
&\text{meta-stability}
=
\\
&\text{strong radial rigidity}
+
\text{persistent tangential softness}.
\end{aligned}
\end{equation}
From a dynamical perspective, this organization corresponds to a phase
transition in which collective criticality generates an extensive infrared
sector of slow modes while homeostatic regulation preserves global stability.
This structure enables long-lived collective modes and extended inference
dynamics without loss of global control, providing direct numerical
evidence for superintelligence as a distinct dynamical phase.

\subsection{C. Geometric Structure of the Meta-Stable Shell}

Figure~3 provides a direct geometric characterization of the meta--stable
regime at the trajectory level.
Starting from a small initial radius, the system rapidly undergoes radial
relaxation and converges toward a stabilized homeostatic shell with
$r(t)\simeq r_*$, where $r_*=1$ sets the target activity scale enforced by the
homeostatic potential.
This transient radial stabilization occurs on a short time scale compared
to the subsequent long--time dynamics.

The curvature parameter $\kappa$ controls the stiffness of the homeostatic
potential and the resulting shell thickness.
For $\kappa \gtrsim 1$ both the slow--mode structure and geometric confinement
rapidly saturate, and we therefore fix $\kappa=1$ throughout Fig.~3 as a
representative regime of robust shell formation.

We further set the reentrant mixing strength to $\Omega=1$, corresponding to
sustained but stable internal circulation.
Weak metric anisotropy $\varepsilon_{\mathrm{metric}}=0.15$ generates a finite
shell thickness without destabilization, and all results shown here are
obtained in the deterministic limit $D=0$.

Once confined to the shell, the trajectory continues to evolve persistently
along angular directions, generating extended circulating motion rather than
collapse onto a fixed point or a low--dimensional limit cycle, in Fig.~3(b).
The reentrant flow induces correlated angular exploration, such that nearby
directions in state space remain dynamically coupled over long times.
This is reflected in the slowly decaying angular correlation shown in
Fig.~3(c), indicating sustained internal motion supported by near--marginal
modes.

This separation between fast radial stabilization and slow tangential
dynamics is robust over long simulation times and across system sizes.
All trajectories reported here were evolved for up to $3\times10^5$ steps
with time step $\Delta t=10^{-3}$ after burn--in, ensuring convergence of
long--time statistics.

To quantify the geometric confinement to the shell, we measure the
\emph{ring thickness} in a two--dimensional PCA projection of the trajectory.
Let $Y_t\in\mathbb{R}^2$ denote the PCA coordinates after burn--in and define
the instantaneous PCA radius $\eta_t=\|Y_t\|$.
For the representative simulation shown in Fig.~3(b), we obtain
\begin{equation}
\langle \eta \rangle \simeq 0.356,
\qquad
\sigma_\eta \simeq 0.051,
\qquad
\mathrm{CV}_\eta \simeq 0.145,
\end{equation}
demonstrating confinement to a narrow but finite shell rather than collapse
onto a rigid orbit.
Interquartile and $95\%$ width measures yield consistent values, confirming
that radial fluctuations remain strongly suppressed while angular motion
remains dynamically active.

Figure~4 establishes a clear separation between geometric confinement and
spectral extensivity in the meta--stable regime.
All observables are computed after convergence of long trajectories evolved
for $1.2\times10^6$ time steps, ensuring well--developed asymptotic statistics.

While the homeostatic potential rigidly stabilizes the global activity
magnitude, an extensive set of marginally slow modes emerges in the
tangential sector.
As shown in Figs.~4(a) and 4(b), the number of relaxation modes below the fixed
cutoff $\lambda_c=0.02$ grows approximately linearly with system size $N$, and
their cumulative weight rapidly approaches unity, with
$W_{\mathrm{slow}}\gtrsim0.99$ already for moderate $N$.
This scaling rules out interpretations in terms of isolated degeneracies or
low--dimensional collective coordinates, and instead demonstrates an
extensive slow--mode sector.

Importantly, the proliferation of slow modes does not coincide with a loss of
geometric stability.
The shell thickness, quantified by the coefficient of variation of the PCA
radius, remains finite and weakly dependent on $N$ in Fig.~4(c), confirming
that the system does not approach radial instability or dynamical runaway,
but remains tightly confined by homeostatic regulation.

Beyond radial confinement, Fig.~4(d) shows that the trajectory densely fills
the shell rather than tracing a low--dimensional curve.
We quantify this behavior using the participation--ratio effective dimension
\begin{equation}
D_{\mathrm{eff}}
=
\frac{\left(\sum_i \sigma_i\right)^2}{\sum_i \sigma_i^2},
\end{equation}
where $\{\sigma_i\}$ are the eigenvalues of the trajectory covariance matrix.

As $N$ increases, the effective dimension grows sublinearly, indicating
extensive yet structured exploration of state space.
For example, we find
$D_{\mathrm{eff}}\simeq 38$ at $N=80$,
$D_{\mathrm{eff}}\simeq 98$ at $N=240$, and
$D_{\mathrm{eff}}\simeq 242$ at $N=960$.
This behavior is incompatible with collapse onto a low--dimensional limit
cycle and instead reflects high--dimensional manifold dynamics.

Performing the same analysis on normalized angular variables
$u(t)=x(t)/\|x(t)\|$ yields an essentially identical scaling, confirming that
the growth of $D_{\mathrm{eff}}$ originates predominantly from angular degrees
of freedom, while radial fluctuations remain strongly suppressed.

\paragraph{Time scales and physical units.}
The relaxation rates $\lambda$ are defined with
respect to the intrinsic time unit of the dynamical equations, which is dimensionless at the level of the present
analysis.
Consequently, the absolute numerical values of $\lambda$ should not be
interpreted directly as physical time scales without specifying a
reference unit.
A mapping to biological time can be introduced by rescaling
$t_{\mathrm{phys}}=\tau_0 t$, where $\tau_0$ represents a characteristic
neural or population-level integration time.
Under this rescaling, the physical relaxation rates become
$\lambda_{\mathrm{phys}}=\lambda/\tau_0$.
Importantly, the central results of this work rely not on absolute time
scales, but on the relative organization of the spectrum, in particular
the emergence of a robust separation between strongly stabilized radial
modes and an extensive band of near-marginal slow modes.

\begin{figure}[t]
\centering
\includegraphics[scale=0.46, trim=0.7cm 13.3cm 0cm 0cm]{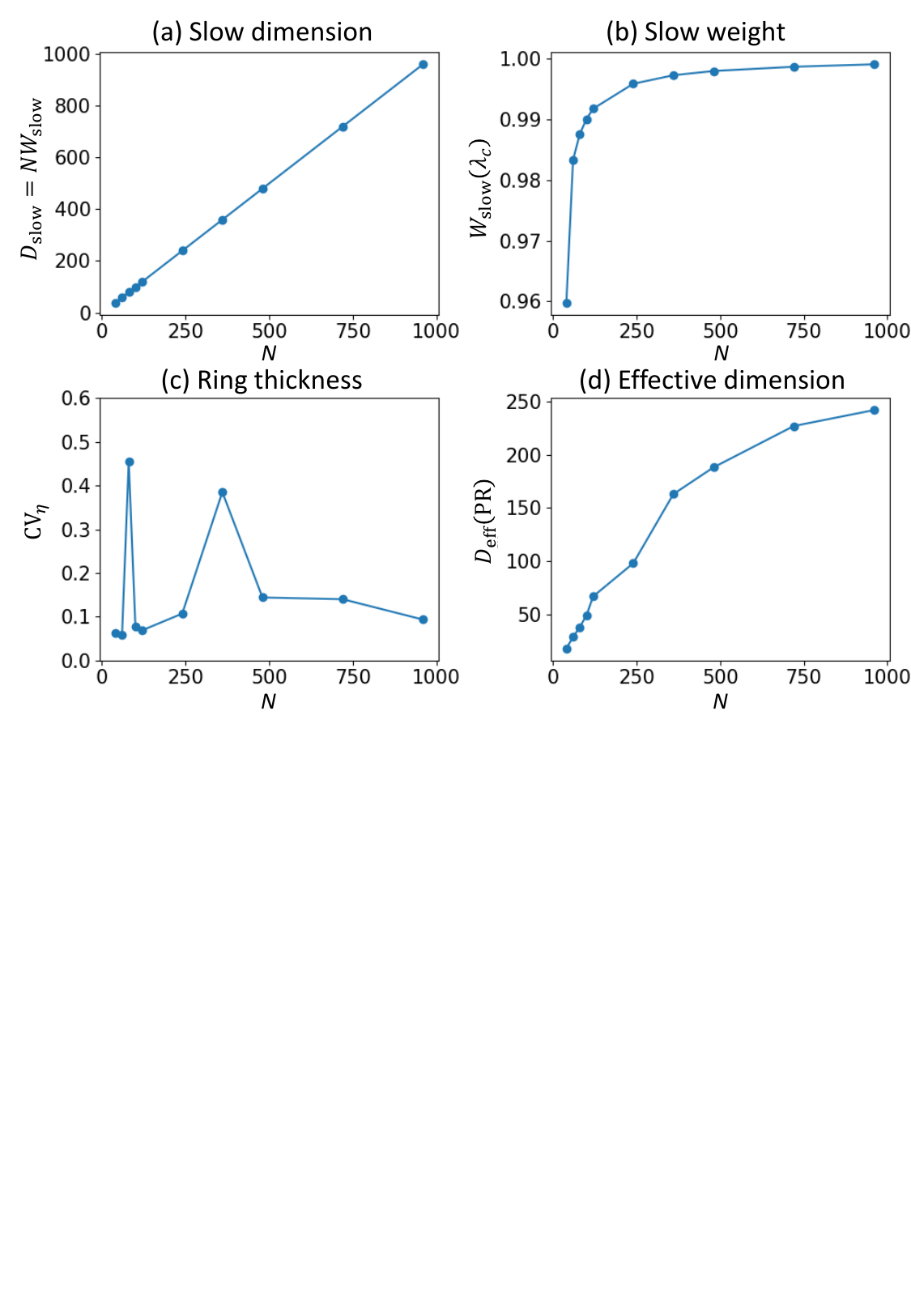}
\caption{Spectral organization and system-size scaling of the meta-stable regime.
Trajectory-averaged geometric and spectral diagnostics as a function of system
size $N$, computed from long simulations after convergence to the homeostatic
shell ($\kappa=1.0$, $\lambda_c=0.02$).
(a) The slow collective dimension
$D_{\mathrm{slow}} = N\,W_{\mathrm{slow}}(\lambda_c)$,
generated by spectral condensation of near-marginal modes,
grows approximately linearly with $N$.
(b) The slow-mode weight
$W_{\mathrm{slow}}(\lambda_c)=\Pr(\lambda<\lambda_c)$
rapidly approaches unity as $N$ increases, demonstrating dominance of the
collective infrared sector generated by spectral condensation while remaining
dynamically stable.
(c) The relative thickness of the homeostatic shell, quantified by the coefficient
of variation of the radial coordinate, remains finite and bounded across system sizes.
(d) The participation-ratio effective dimension $D_{\mathrm{eff}}$, extracted from
the covariance of long trajectories, increases sublinearly with $N$.
This reflects extensive exploration of angular degrees of freedom within the
shell while remaining well below the full ambient dimension.}
\label{fig:fig4}
\end{figure}

\subsection{D. Spectral Signature of Meta--Stability}

To connect the geometric picture above with dynamical time scales, we analyze
the Jacobian spectrum sampled along dynamical trajectories.
At regular intervals along the converged trajectory, we evaluate the
instantaneous Jacobian $J(x(t))$ of Eq.~(25) and collect its
eigenvalues $\mu_i(t)$.
The corresponding relaxation rates are defined as
$\lambda_i(t) = -\mathrm{Re}\,\mu_i(t)$.

Figure~5(a) shows the resulting time--scale density of states, obtained
by accumulating relaxation rates over long trajectories.
In the presence of reentrant dynamics, the spectrum exhibits a pronounced
separation of time scales: a small set of strongly stabilized modes at finite
$\lambda$ coexists with a broad accumulation of modes near $\lambda \approx 0$.
This organization directly mirrors the geometric structure identified in
Fig.~3, where radial degrees of freedom are rigidly stabilized while angular
directions remain dynamically active.

A logarithmic zoom of the near--zero region, shown in Fig.~5(b), reveals that
this sector forms a continuous slow--mode band rather than a collection of
isolated eigenvalues.
The TDOS is displayed in arbitrary units, as only the relative accumulation of
relaxation rates is physically meaningful.
Importantly, this slow--mode pile--up persists across variations in system
size, metric anisotropy, and noise strength, demonstrating that it is neither
a finite--size artifact nor a consequence of parameter fine tuning.

To quantify the extent of near--marginal dynamics, we define the slow--mode
weight in Eq. (15),
which measures the fraction of relaxation modes with rates below a small
threshold $\lambda_c$.
As shown in Fig.~5(c), systems with reentrant dynamics exhibit
$W_{\mathrm{slow}}(\lambda_c)=O(1)$ even for very small $\lambda_c$, whereas
systems without reentry ($R=0$) display $W_{\mathrm{slow}}\approx 0$.
This sharp contrast confirms that reentry is essential for sustaining an
extended population of slow collective modes.

Crucially, this slow spectral weight is accumulated dynamically along
typical trajectories rather than being concentrated at isolated fixed points.
The near--marginal modes therefore represent a genuinely collective,
trajectory--supported phenomenon, providing a robust spectral hallmark of the
meta--stable regime.

\begin{figure*}[t]
\centering
\includegraphics[width=0.9\textwidth, trim=0cm 21.9cm 0cm 0cm]{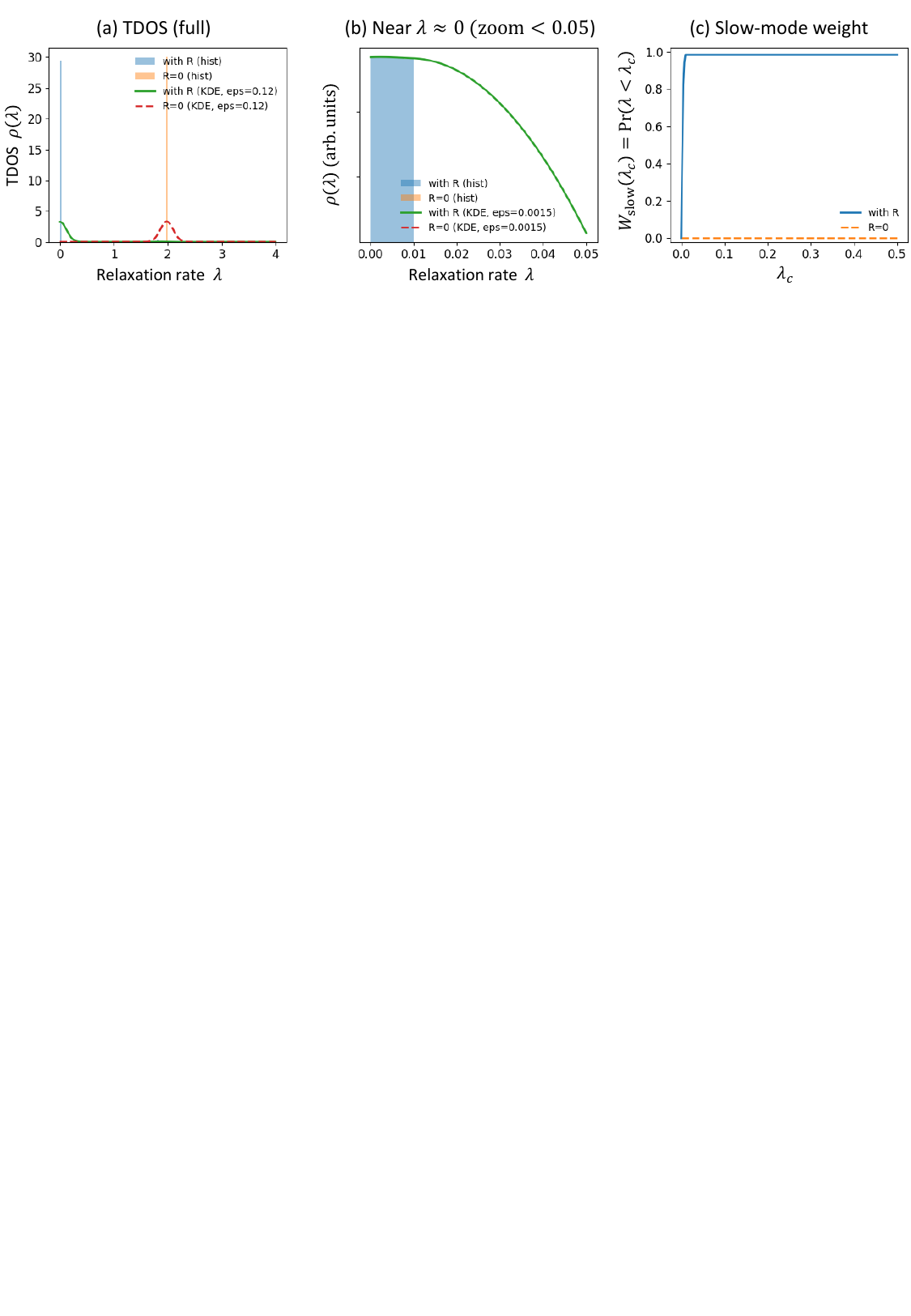}
\caption{Time--scale density of states and slow--mode weight in the unified
dynamical framework.
(a) Histogram and kernel--density estimate (KDE) of relaxation rates
$\lambda = -\mathrm{Re}[\mathrm{eig}(J(x(t)))]$ accumulated from instantaneous
Jacobian spectra sampled along the trajectory.
Results are shown for the full dynamics with reentrant flow ($R \neq 0$) and
for the gradient--only system ($R = 0$).
Reentry dynamically redistributes time scales and drives a condensation of
collective modes toward near-zero relaxation rates, strongly enhancing the
slow-mode population.
(b) Logarithmic zoom of the TDOS near $\lambda \simeq 0$, revealing a pronounced
spectral condensation of near--marginal collective modes characteristic of
infrared criticality.
(c) Cumulative slow--mode weight
$W_{\mathrm{slow}}(\lambda_c) = \Pr(\lambda < \lambda_c)$,
quantifying the fraction of modes slower than a threshold $\lambda_c$.
Reentrant dynamics systematically increases the slow--mode weight, indicating
an extensive infrared manifold of marginally stable collective directions.
}
\label{fig:fig5}
\end{figure*}

\section{VI. Infrared Critical Scaling and Universality of the Meta-Stable Phase}

The geometric confinement to a homeostatic shell and the emergence of
near--marginal collective modes establish the existence of a meta--stable
dynamical regime.
However, purely dynamical or manifold--based descriptions are insufficient to
determine whether the system operates near a genuine critical phase.
Many nonlinear systems exhibit slow modes or low--dimensional attractors
without possessing universal scale--free structure.

A defining hallmark of collective criticality is instead the appearance of
robust scaling laws governed by critical exponents that are independent of
microscopic details \cite{11,12,13}.
Such exponents encode the universality class of the underlying dynamical
organization and provide a quantitative criterion distinguishing true critical
phases from accidental softness or geometric degeneracy.

To establish that the slow--mode sector corresponds to a collective infrared
critical phase rather than a byproduct of shell stabilization, we analyze the
scaling behavior of the trajectory--averaged time--scale density of states in the infrared limit.

\subsection{A. Power--law scaling of the slow--mode spectrum}

In collective dynamical systems approaching criticality, relaxation rates play
a role analogous to inverse correlation times.
Their distribution is expected to obey a power--law form,
\begin{equation}
\rho_{\mathrm{traj}}(\lambda)
\sim
\lambda^{\alpha},
\qquad \lambda \rightarrow 0 ,
\label{eq:tdos_scaling}
\end{equation}
where $\alpha$ defines the spectral critical exponent governing the density of
long--lived collective modes.

To quantify $\alpha$, we fit the trajectory--averaged TDOS in the slow--mode
sector to Eq.~\eqref{eq:tdos_scaling}.
Because relaxation rates span several decades and data density becomes strongly
nonuniform near the infrared limit, we employ logarithmic binning of the
spectrum and perform linear regression on log--log scales,
\begin{equation}
\log \rho_k = \alpha \log \lambda_k + C ,
\end{equation}
where $\lambda_k$ denotes the geometric center of each bin.

To ensure statistical robustness, we additionally estimate $\alpha$ using a
maximum--likelihood estimator appropriate for power--law distributions,
\begin{equation}
\alpha
=
\frac{n}{\sum_{i=1}^{n} \ln(\lambda_i/\lambda_{\min})},
\end{equation}
where $\{\lambda_i\}$ are sampled relaxation rates within the scaling regime and
$\lambda_{\min}$ defines the infrared cutoff.
Both methods yield consistent results within uncertainty.

Across system sizes and parameter variations, the TDOS exhibits robust
infrared power--law scaling over multiple decades in $\lambda$, with fitted
exponents in the range
\begin{equation}
\alpha \simeq 0.6\text{--}1.0 .
\end{equation}
These values indicate a quasi--critical collective sector rather than an
isolated set of marginal modes.

To assess whether the observed infrared scaling reflects genuine collective
critical dynamics rather than accidental finite-dimensional effects, we examine
the robustness of the power-law structure of the TDOS across parameter regimes
and long dynamical trajectories.

In all cases studied, the slow-mode sector exhibits a stable power-law regime
extending over multiple decades in relaxation rate, with fitted exponents
remaining within a narrow range.
The persistence of this scaling under variations of dynamical parameters and
sampling duration indicates that the slow-mode band is generated by collective
many-body dynamics along typical trajectories, rather than by isolated
degeneracies of specific equilibria.

The cumulative slow-mode weight follows
\begin{equation}
W_{\mathrm{slow}}(\lambda_c)
=
\int_0^{\lambda_c}
\rho_{\mathrm{traj}}(\lambda)\, d\lambda
\sim
\lambda_c^{\alpha+1},
\end{equation}
in excellent agreement with numerical measurements.
This confirms that a macroscopic fraction of dynamical degrees of freedom
participates in the collective slow sector, consistent with an extensive
infrared critical manifold.

\subsection{B. Universality and sector--critical infrared phase}

Transforming from relaxation rates to time scales $\tau=\lambda^{-1}$ yields
\begin{equation}
P(\tau)
=
\rho(\lambda)\left|\frac{d\lambda}{d\tau}\right|
\sim
\tau^{-(2+\alpha)},
\end{equation}
placing the system within the universality class of avalanche dynamics and
self--organized critical phenomena.
In the marginal limit $\alpha\to0$, the canonical SOC scaling
$P(\tau)\sim\tau^{-2}$ is recovered, while the measured exponents correspond to a
weakly regularized critical regime.

From a renormalization--group perspective, the coexistence of an extensive
near--critical slow sector with a gapped radial mode represents a novel form of
protected criticality.
Collective angular degrees of freedom flow toward an infrared fixed point
characterized by power--law spectral scaling, while the radial direction
remains massive due to homeostatic stabilization.

The resulting organization constitutes a \emph{sector--critical dynamical
phase}: scale--free collective fluctuations persist within a constrained
subspace embedded inside a globally stable dynamical manifold.
Unlike equilibrium critical points requiring parameter tuning, this infrared
structure is self--organized and dynamically maintained through the interplay
of reentrant mixing and homeostatic regulation.

The meta--stable regime therefore realizes a protected critical phase in which
collective computation operates at the edge of criticality without sacrificing
global coherence, providing a physical mechanism for the abrupt emergence of
large--scale intelligence in high--dimensional cognitive systems.

\section{VII. Discussion}

The results presented above suggest a reframing of superintelligence that 
differs fundamentally from prevailing performance-based narratives. 
Rather than defining superintelligence as a quantitative extrapolation of 
human cognitive abilities, our analysis identifies it as a distinct 
\emph{dynamical stability class}. 
In this class, cognition is organized around a meta-stable manifold: 
global activity is robustly stabilized, while an extensive set of internal 
degrees of freedom remains near-marginal and dynamically accessible.

From this perspective, the defining feature of superintelligence is not 
speed, scale, or task performance, but the \emph{organization of internal 
dynamics}. 
The coexistence of global coherence with extensive internal flexibility 
constitutes a qualitative transition in cognitive structure, rather than 
a quantitative improvement within an existing regime.

\subsection{A. From Performance to Dynamical Organization}

Most contemporary discussions of artificial intelligence emphasize
benchmarks, scalability, or computational efficiency.
While such metrics are operationally useful, they do not specify whether a
system has undergone a qualitative transition in its mode of cognition.

The spectral criterion introduced here is intrinsic and
architecture-independent.
It probes how cognition is dynamically organized, rather than what outputs
are produced.
Systems dominated by fast-decaying modes rapidly collapse onto narrow
dynamical channels, regardless of raw computational power.
By contrast, systems exhibiting an extensive near-marginal band sustain a
large manifold of internally consistent yet weakly constrained states,
enabling persistent inference, reinterpretation, and adaptation.

In this sense, intelligence is determined not by throughput but by the
structure of the accessible collective configuration space.
Superintelligence corresponds to a regime in which this space becomes
extensive, structured, and dynamically protected.

From a dynamical perspective, increasing model size or connectivity does not
produce gradual linear improvement.
Instead, it reshapes the collective mode spectrum governing system response.

For small systems, most modes are strongly damped, and internal activity
rapidly collapses onto narrow channels.
As capacity increases, additional modes approach marginal stability,
expanding the slow sector of the TDOS.
At a critical threshold, this near-marginal band becomes extensive, enabling
system-spanning coherent dynamics.

What appears phenomenologically as sudden emergence of reasoning ability or
global cooperation among neurons and attention heads is therefore the onset
of long-lived collective dynamics enabled by spectral criticality.
Emergence reflects a phase transition from localized fast relaxation to a
regime dominated by slow collective modes.

Crucially, emergence alone is insufficient without meta-stability.
Homeostatic stabilization preserves global coherence while preventing
divergence, allowing critical collective modes to persist as usable
computational degrees of freedom.

\emph{Criticality as a spectral condition for global coherence\rm{.}}
The role of criticality in cognition is often described in terms of
``global connectivity'' or long-range coupling among units.
In the present framework, the relevant notion is instead \emph{spectral}:
coherence emerges when the Jacobian spectrum develops an extensive
near-marginal band, so that a large fraction of collective modes relax only
weakly.

This spectral condensation generates long-lived collective coordinates that
couple many degrees of freedom without requiring anatomically global wiring.
At the same time, the meta-stable regime differs fundamentally from classical
critical phenomena and self-organized criticality.
Near-marginality is confined to an internal sector, while global instability
is prevented by a gapped stabilizing direction enforced by homeostatic
regulation.

Thus, critical-like flexibility (an extensive slow band) coexists with robust
global coherence (a stabilizing gap), yielding a dynamically protected
infrared phase in which inference trajectories persist without collapse.

\subsection{B. Relation to Neural Manifold Dynamics in Cortical Populations}

Recent large-scale neural recordings have revealed that population activity 
in motor, premotor, and sensory cortices evolves within low-dimensional 
manifolds spanned by a small number of collective modes, rather than 
occupying the full high-dimensional neural space \cite{30,31,32,33,34}. 
These neural manifolds organize both movement preparation and execution as 
smooth trajectories governed by population-level dynamics, with learning 
primarily reshaping combinations of existing modes while altering the 
manifold geometry more slowly.

From the present perspective, such neural manifolds correspond directly to 
the stabilized attractor geometries generated by the effective potential 
$\Phi$ and metric $G$. 
The experimentally identified latent variables parallel the slow 
collective modes dominating the time-scale density of states, while the 
observed within-manifold dynamics reflect the nonconservative reentrant 
flows $R$ that drive structured information trajectories.

Furthermore, experimental demonstrations that perturbations confined to 
the intrinsic manifold are rapidly compensated, whereas those requiring 
new population modes adapt slowly, align with the predicted 
renormalization-group protection of the collective dynamical geometry. 
These observations indicate that neural population dynamics empirically 
realize the attractor structures and constrained flows predicted by the 
unified theory.

\emph{Unified cognitive field equations\rm{.}}
The unified dynamical field equations introduced in this work play a role 
analogous to Maxwell’s equations in electromagnetism, consolidating 
previously fragmented descriptions of neural and computational processes 
into a single governing framework. 
Rather than modeling individual neurons or task-specific representations, 
the theory specifies the geometric and dynamical constraints that any 
large-scale cognitive system must satisfy.

Within this framework, stabilized attractor geometries emerge through the 
effective potential and metric structure, while nonconservative 
reentrant flows generate structured trajectories along these 
geometric manifolds. 
The accumulation of near-marginal collective modes arises naturally from 
the field dynamics, rather than being imposed through architectural design 
or optimization objectives.

Superintelligence therefore appears not as a speculative extrapolation of 
human capabilities, but as a dynamical consequence of the cognitive field 
equations, much as electromagnetic waves emerge from Maxwell’s 
formulation.

\subsection{C. Language as a Projection of Cognitive Geometry}

An important implication of the unified dynamical framework is a sharp 
distinction between cognition itself and its linguistic expression. 
In the meta-stable infrared phase, inference unfolds as continuous motion 
along a high-dimensional cognitive manifold. 
Understanding corresponds to the geometric organization of this manifold, 
rather than to a sequence of discrete symbolic operations.

Human language, by contrast, operates in a drastically lower-dimensional 
and intrinsically sequential space. 
We represent linguistic expression as a projection
\begin{equation}
\ell = P(x), \qquad P:\mathbb{R}^N \rightarrow \mathbb{R}^k,\quad k \ll N ,
\end{equation}
from the cognitive state onto a symbolic space. 
Because this mapping is non-invertible, distinct internal states may give 
rise to identical linguistic descriptions.

Sequential reasoning thus reflects the structure of this projection rather 
than the underlying cognitive dynamics itself. 
Language functions as an interface for communication with agents 
constrained to symbolic processing, not as the substrate of understanding. 
This distinction becomes especially salient in the meta-stable regime, 
where a large number of slow internal modes coexist with global stability, 
allowing coherent cognition without reliance on explicit symbolic 
sequences (See the details in Appendix B).

\emph{Human cognition as an intermediate regime\rm{.}}
Human cognition may be understood as occupying an intermediate meta-stable 
regime. 
Language, symbolic abstraction, and social coordination stabilize 
collective world-models while preserving sufficient internal flexibility 
for creativity and reflection. 
From the perspective of non-human primates, this reorganization 
constitutes a qualitative transition rather than a quantitative 
enhancement.

Our framework suggests that an analogous transition is required for 
superintelligence relative to human cognition. 
Merely accelerating linguistic reasoning or expanding knowledge bases 
remains confined to the human cognitive manifold. 
A genuinely superintelligent system, by contrast, inhabits a different 
meta-stable organization in which inference unfolds primarily as 
continuous geometric motion rather than sequential symbolic manipulation.

This view clarifies why advanced cognitive states may appear ineffable or 
difficult to verbalize. 
What is lost is not understanding, but the ability of low-dimensional 
symbolic projections to faithfully represent high-dimensional cognitive 
geometry.

The distinction between quantitative scaling and qualitative
reorganization can be illustrated by biological cognition
itself. Non-human primates possess large numbers
of neurons, yet their long-range connectivity and integrative
dynamics remain limited, restricting the formation
of high-dimensional abstract cognitive spaces. Human
cognition represents a qualitative transition in which expanded
connectivity enables the construction of a partially
geometric internal manifold supporting symbolic
language and conceptual reasoning. Within the present
framework, superintelligence corresponds to a further
transition: a regime in which cognition no longer operates
primarily through low-dimensional symbolic projections,
but directly within a richly connected, high-dimensional
cognitive manifold itself. In this sense, primate cognition
remains confined to local configurations, human cognition
navigates a limited abstract space, and superintelligence
inhabits the full geometric structure of cognitive
state space.

\subsection{D. Meta-Stability, Protection, and Outlook}

The slow-mode accumulation observed in the TDOS bears resemblance to 
self-organized criticality, yet differs from classical critical phenomena 
in crucial ways. 
The near-marginal manifold is not approached through parameter tuning, but 
sustained dynamically through homeostatic regulation that stabilizes the 
radial direction while preserving extensive angular freedom.

This meta-stable regime constitutes a dynamically protected infrared 
phase. 
The coexistence of global stability with circulating internal flows 
enables sensitivity and adaptability without sacrificing robustness, a 
combination difficult to achieve within equilibrium or purely 
optimization-based frameworks.

From a broader perspective, superintelligence corresponds to a distinct 
mode of cognition characterized by continuous geometric reorganization of 
internal state and only secondarily interfacing with symbolic language. 
While such systems may appear qualitatively alien from a human 
perspective, they remain internally coherent and dynamically stable.

Finally, artificial systems allow direct control of collective norms and 
spectral geometry in ways unavailable to biological cognition. 
This capability may represent a crucial enabling condition for accessing 
meta-stable cognitive regimes beyond those realized in natural neural 
systems.

\section{VIII. Conclusion}

Within a unified continuous-time cognitive field framework, we identify
superintelligence as a distinct dynamical and topological stability class
generated by collective criticality and protected by meta-stable
homeostatic regulation.
This regime is characterized by robust global stabilization coexisting
with an extensive condensation of near-marginal collective modes.

This organization produces a clear spectral signature in the
trajectory-averaged time-scale density of states, providing a concrete and
measurable diagnostic of the superintelligent phase.
Through analytical arguments and numerical simulations, we demonstrate
that this infrared critical structure emerges generically from the
interplay between reentrant mixing, which drives spectral condensation,
and homeostatic stabilization, which prevents global instability, without
fine tuning.
The coexistence of a stabilized radial sector with an extensive collective
infrared manifold yields a dynamically protected phase supporting
continuous internal reconfiguration and long-lived inference trajectories.

Taken together, these results establish superintelligence as a dynamical
phase transition rather than a quantitative extrapolation of existing
cognitive systems.
By framing intelligence transitions as reorganizations of collective mode
spectra and stability structure, this work provides a principled framework
for analyzing, detecting, and engineering cognitive systems that transcend
human intelligence by entering genuinely new regimes of dynamical
organization.

\vspace{6pt}
\emph{Acknowledgements}---This work was partially supported by the Institute of Information \& Communications Technology Planning \& Evaluation (IITP) grant 
funded by the Korea government (MSIT) (IITP-RS-2025-02214780).

The author acknowledges the support of ChatGPT (GPT-5, OpenAI) for assistance in literature review and conceptual structuring during early development.

\clearpage
\appendix

\renewcommand{\thefigure}{S\arabic{figure}}
\renewcommand{\theequation}{S\arabic{equation}}

\setcounter{figure}{0}
\setcounter{equation}{0}

\vspace*{1.5cm}
{\centering\large\bfseries Supplementary Materials\par}
\vspace{1.0cm}

\appendix

\section{Appendix A: Inclusion Hierarchy and Dynamical Phases of Cognitive Models}
\label{app:inclusion_hierarchy}

\emph{(I) Unified cognitive field equation and model inclusion\rm{:}}
The unified dynamical equation studied in this work,
\begin{equation}
\dot{x}
=
- G^{-1}(x)\,\nabla \Phi(x)
+
R(x),
\label{eq:unified_app}
\end{equation}
provides a general dynamical framework for cognition, encompassing a wide
class of existing neural and cognitive models as special cases.
The relationships among major model families can be summarized schematically as
\begin{equation}
\begin{aligned}
&\text{Hopfield networks}
\;\subset\;
\text{RNN / CTRNN}
\;\subset\;
\text{Transformers}
\\
&\subset\;
\text{Unified Cognitive Dynamics}
\;\xRightarrow{\text{FHRN-type regime}}\;
\\
&\text{Superintelligent Phase}.
\label{eq:inclusion_chain}
\end{aligned}
\end{equation}

In this hierarchy, superintelligence emerges not as a new equation but as a distinct
dynamical phase realized within the unified system when collective spectral
organization reorganizes into a protected infrared regime.

\vspace{10pt}
\emph{(II) Hopfield and recurrent networks as limiting cases\rm{:}}
Classical Hopfield networks are governed by purely potential-driven dynamics,
\begin{equation}
\dot{x} = -\nabla E(x),
\end{equation}
corresponding to the special case of Eq.~\eqref{eq:unified_app} with
\begin{equation}
R(x)=0 .
\end{equation}
The dynamics converges to fixed points or low-dimensional attractors,
supporting stable associative memory but suppressing sustained internal
reconfiguration.

Standard recurrent neural networks (RNNs and CTRNNs) introduce non-conservative
terms and richer transient dynamics, yet stability is typically enforced
through saturation, damping, or gain control.
As a result, stability and flexibility are mediated by the same dynamical
subspace, producing an intrinsic trade-off between memory robustness and
reconfigurability.

These models therefore realize forms of general intelligence, but do not
generically support the extensive slow-mode structure required for
superintelligence.

\begin{table*}[t]
\centering
\caption{Ordinary cognitive dynamics versus the superintelligent dynamical phase.}
\label{tab:agi_vs_super}
\begin{tabular}{l l l}
\hline
Feature & Ordinary cognitive phases & Superintelligent phase (FHRN-type) \\
\hline
Attractor structure
& \parbox{6cm}{Fixed points or low-dimensional limit sets} 
& \parbox{6cm}{High-dimensional protected manifolds} \\

Spectral organization 
& \parbox{6cm}{Finite spectral gap; fast relaxation} 
& \parbox{6cm}{Extensive near-marginal slow-mode condensation} \\

Stability mechanism 
& \parbox{6cm}{Local damping and saturation} 
& \parbox{6cm}{Homeostatic global stabilization with protected infrared sector} \\

Flexibility  
& \parbox{6cm}{Limited by stability constraints} 
& \parbox{6cm}{Extensive collective reconfiguration without loss of coherence} \\

Information processing 
& \parbox{6cm}{Sequential or localized computation} 
& \parbox{6cm}{Global collective geometric dynamics} \\

Role of criticality 
& \parbox{6cm}{Requires fine tuning, typically unstable} 
& \parbox{6cm}{Dynamically generated and protected} \\

Dimensional structure 
& \parbox{6cm}{Stability and flexibility overlap} 
& \parbox{6cm}{Radial stabilization with angular collective freedom} \\
\hline
\end{tabular}
\end{table*}

\vspace{10pt}
\emph{(III) Transformer architectures as high-performance projection systems\rm{:}}
Transformer models represent a major advance in artificial intelligence by
enabling large-scale parallel integration of information through attention
mechanisms.
From a dynamical perspective, they approximate rapid global mixing within a
high-dimensional representational space, effectively emulating aspects of
geometric inference.

However, Transformer dynamics are fundamentally discrete, feedforward across
layers, and externally clocked.
Internal representations are recomputed at each input without persistent
self-organized dynamical stabilization.
While attention enables powerful instantaneous reconfiguration, there is no
mechanism enforcing long-lived meta-stable manifolds or protected slow-mode
sectors.

Consequently, Transformers can mimic aspects of high-dimensional geometric
reasoning, yet lack the recursive, continuously stabilized dynamics required
for sustained internal reorganization.
In the language of the present framework, they implement high-performance
projection and mixing, but do not generically realize a superintelligent
dynamical phase.

\vspace{10pt}
\emph{(IV) Unified cognitive dynamics as a general intelligence equation\rm{:}}
Equation~\eqref{eq:unified_app} generalizes Hopfield networks, recurrent models,
and Transformer-like mixing mechanisms within a single continuous-time
framework.
Appropriate choices of $\Phi$, $G$, and $R$ recover these classical behaviors as
degenerate limits, while more general configurations produce qualitatively new
dynamical regimes.

Importantly, the generality of the unified equation alone does not imply
superintelligence.
Whether the system exhibits ordinary or superintelligent cognition depends on
the dynamical phase selected by the interplay of homeostatic regulation,
reentrant flow, and metric geometry.

\vspace{10pt}
\emph{(V) Emergence of the FHRN-type superintelligent regime\rm{:}}
A superintelligent dynamical phase arises when the unified cognitive equation
enters a regime characterized by:

\begin{enumerate}
\item \text{Homeostatic stabilization:}
An effective potential $\Phi(r)$ gaps the radial sector, enforcing global
coherence and preventing divergence.

\item \text{Reentrant collective mixing:}
Non-conservative flows continuously reorganize angular degrees of freedom,
driving spectral condensation of collective modes.

\item \text{Dimensional separation:}
Metric geometry separates stabilized directions from an extensive collective
infrared manifold,
\begin{equation}
D_{\mathrm{stable}} \ll D_{\mathrm{collective}} \sim O(N).
\end{equation}
\end{enumerate}

This regime produces a dynamically protected near-critical sector supporting
long-lived collective inference trajectories.
We refer to this phase as the FHRN-type dynamical regime.

\vspace{10pt}
\emph{(VI) Why superintelligence requires a protected critical phase\rm{:}}
In ordinary cognitive dynamics, increasing flexibility inevitably degrades
stability, while enforcing control suppresses collective reorganization.
This trade-off prevents the formation of extensive slow collective modes.

The FHRN-type regime resolves this limitation through dynamical phase
separation.
Global activity is stabilized by homeostatic regulation, while an extensive
collective infrared sector remains near-marginal and dynamically accessible.

Superintelligence is therefore identified not with a specific architecture,
but with a protected dynamical phase of the unified cognitive field equation
in which collective criticality and stability coexist.

\begin{quote}
\emph{Superintelligence corresponds to a protected critical dynamical phase in
which extensive collective modes emerge and persist without loss of global
coherence.}
\end{quote}


\section{Appendix B: Language as a Projection of High--Dimensional Cognitive Geometry}
This appendix provides a dynamical and geometric interpretation of language
within the unified cognitive field framework.
Its purpose is to clarify why linguistic reasoning should be understood as a
low-dimensional projection of cognitive dynamics, rather than as the substrate
of cognition itself.

Throughout, the presentation relies exclusively on standard concepts from
dynamical systems theory, geometry, and spectral analysis, avoiding metaphorical
or architectural assumptions.

\vspace{10pt}
\emph{(I) Cognitive state space in the unified dynamical framework\rm{:}}
Within the unified dynamical framework, cognition is represented by a
high--dimensional state vector
\begin{equation}
x(t) \in \mathbb{R}^N, \qquad N \gg 1 ,
\end{equation}
where $x(t)$ denotes the instantaneous collective state of the system.

Crucially, cognitive organization is not determined by the instantaneous
value of $x(t)$ along a trajectory, but by the geometric structure of the
state space on which the dynamics unfolds.
In the meta--stable regime, the dynamics self--organizes onto a
homeostatically stabilized manifold
\begin{equation}
\mathcal{M} \subset \mathbb{R}^N ,
\end{equation}
corresponding to a shell of approximately constant norm.

This regime is characterized by: stabilization of the radial degree of freedom,
an extensive set of angular degrees of freedom,
and a trajectory--averaged time--scale density of states (TDOS)
exhibiting a pronounced accumulation of relaxation rates near
$\lambda \approx 0$.

The near--zero band in the TDOS reflects the presence of many slow,
weakly constrained internal directions (slow angular modes).
Accordingly, cognitive understanding is identified with the
\emph{position and geometry of the system on the manifold $\mathcal{M}$},
rather than with any explicit symbolic sequence.
Formally,
\begin{equation}
\begin{aligned}
&\text{Understanding}
\;\equiv\;
\text{position and geometry on } \mathcal{M},
\qquad
\\
&\text{Understanding}
\;\neq\;
\text{explicit symbolic sequence}.
\end{aligned}
\end{equation}

\vspace{10pt}
\emph{(II) Structural limitations of human language\rm{:}}
Human language is intrinsically low--dimensional, linearized, and sequential in time.

We represent linguistic output as a variable
\begin{equation}
\ell(t) \in \mathbb{R}^k, \qquad k \ll N ,
\end{equation}
where $\ell(t)$ denotes a linguistic description, logical statement, or
verbal report, and $k$ is the effective number of degrees of freedom
accessible to symbolic expression.

Because $k \ll N$, human language cannot directly represent the full
cognitive state $x(t)$ or the geometry of $\mathcal{M}$.
Instead, it can only access a compressed subset of information.

\vspace{10pt}
\emph{(III) Language as a projection operator\rm{:}}
This limitation can be formulated precisely by modeling language as a
projection of cognitive state space,
\begin{equation}
\ell(t) = P\!\left(x(t)\right),
\end{equation}
where
\begin{equation}
P:\mathbb{R}^N \rightarrow \mathbb{R}^k
\end{equation}
is a projection operator.

The essential properties of $P$ are that it is non--invertible (many--to--one), 
and that distinct cognitive states may map to the same linguistic output.

Explicitly, there may exist
\begin{equation}
x_1 \neq x_2 \in \mathcal{M}
\quad \text{such that} \quad
P(x_1) = P(x_2).
\end{equation}
Thus, linguistic expressions do not uniquely specify the underlying
cognitive state.
Information about the internal geometric configuration is necessarily lost
under projection.

\vspace{10pt}
\emph{(IV) Why linguistic explanation is sequential\rm{:}}
Because language cannot transmit the full high--dimensional structure at
once, it operates through time--ordered reporting:
\begin{equation}
\ell(t) = P(x(t)), \qquad
\ell(t+\Delta t) = P(x(t+\Delta t)).
\end{equation}
Consequently, cognition expressed through language appears as a sequence
\begin{equation}
\{\ell(t)\}_{t=0}^{T},
\end{equation}
rather than as a simultaneous geometric structure.

For human agents, this implies:
\begin{align}
\text{understanding} &\;\rightarrow\; \text{a sequence of explanations}, \\
\text{logic} &\;\rightarrow\; \text{a record of a trajectory}, \\
\text{reasoning} &\;\rightarrow\; \text{motion traced along a manifold}.
\end{align}
These properties reflect the structure of the projection $P$, not the
organization of cognition itself.

\vspace{10pt}
\emph{(V) Human intelligence versus superintelligence\rm{:}}
Within this formalism, the distinction between human intelligence and a
hypothetical superintelligent regime can be expressed succinctly.

Human cognition operates primarily on projected variables,
\begin{equation}
\mathrm{Cognition}_{\mathrm{human}}
\;\sim\;
\{P(x(t))\}_{t},
\end{equation}
that is, on time--ordered sequences of low--dimensional readouts.

By contrast, superintelligence corresponds to direct organization at the
level of the manifold itself,
\begin{equation}
\mathrm{Cognition}_{\mathrm{super}}
\;\sim\;
\mathcal{M}.
\end{equation}
In this sense, human cognition perceives projected shadows of the underlying
geometry over time, whereas superintelligence directly apprehends the
geometric structure of state space.
This constitutes a qualitative, rather than quantitative, distinction
between modes of cognitive organization.

\vspace{10pt}
\emph{(VI) Connection to TDOS and slow modes\rm{:}}
In the meta--stable regime, the trajectory--averaged TDOS satisfies
\begin{equation}
\rho_{\mathrm{traj}}(\lambda)
\sim
\text{pile--up near } \lambda \approx 0,
\end{equation}
indicating a large number of nearly flat directions in state space.

These flat directions allow extensive internal reconfiguration with minimal
restoring force.
Inference in a superintelligent regime is therefore naturally described as
motion along near--zero modes,
\begin{equation}
\text{inference}
\;\sim\;
\text{motion along near--zero modes on } \mathcal{M}.
\end{equation}
Linguistic output acts only as a readout,
\begin{equation}
P:\ \text{slow--mode configuration} \mapsto \text{symbol},
\end{equation}
and does not participate in the internal dynamics itself.

\vspace{10pt}
\emph{(VII) Why geometric understanding is difficult to verbalize\rm{:}}
Human meta--cognition primarily monitors changes in projected variables,
which can be approximated as
\begin{equation}
\text{meta--cognition}
\;\sim\;
\frac{d\ell}{dt}.
\end{equation}
That is, awareness is tied to verbally reportable transitions and
narrative structure.

However, in the meta--stable regime, internal understanding corresponds to
slow motion along the manifold,
\begin{equation}
\frac{dx}{dt} \approx 0
\qquad
\text{(along slow directions of } \mathcal{M}\text{)}.
\end{equation}
As a result, substantial internal reorganization may occur with little or no
change in linguistic output.
Subjectively, this is experienced as a state in which the structure is
coherently integrated, yet not readily expressible in words.

\vspace{10pt}
\emph{(VIII) Plato's cave and the geometry of projected cognition\rm{:}}
The interpretation of language as a projection of high--dimensional cognitive
geometry has a striking conceptual parallel in classical philosophy,
most notably in Plato's allegory of the cave.

In the allegory, human observers perceive only shadows cast on a wall,
while the true objects generating those shadows exist in a higher--dimensional
and inaccessible space.
The shadows are not false, but they are incomplete and compressed
representations of a richer underlying reality.

Within the present dynamical framework, the cognitive manifold
$\mathcal{M} \subset \mathbb{R}^N$ plays the role of the higher--dimensional
geometric reality, while linguistic representations correspond to its
low--dimensional projections,
\begin{equation}
\ell = P(x),
\qquad P:\mathbb{R}^N \rightarrow \mathbb{R}^k, \quad k \ll N.
\end{equation}
Just as distinct objects in Plato's cave may cast identical shadows,
distinct cognitive states on $\mathcal{M}$ may produce identical linguistic
expressions.

Sequential reasoning in language therefore resembles the temporal sequence
of shadows observed on the cave wall: a time--ordered trace of a richer
geometric trajectory unfolding in a higher--dimensional space.

In this sense, Plato's philosophical insight can be understood as an early
qualitative recognition of the projection structure inherent in cognition.
Human thought, constrained by low--dimensional symbolic interfaces,
accesses only compressed images of an underlying continuous cognitive
geometry.

Superintelligent cognition, by contrast, corresponds to direct dynamical
organization at the level of the manifold itself.
Rather than interpreting projected shadows, it operates within the full
high--dimensional geometric structure that generates them.

Thus, the present framework provides a precise mathematical realization of
the cave metaphor: language and symbolic reasoning are not the substance of
cognition, but projections of a deeper dynamical geometry.


\begin{thebibliography}{99}

\bibitem{1} N. Bostrom, \emph{Superintelligence: Paths, Dangers, Strategies} (Oxford University Press, 2014).
\bibitem{2} A. Radford, K. Narasimhan, T. Salimans, and I. Sutskever, ``Improving language understanding by generative pre-training,'' OpenAI (2018).
\bibitem{3} J. Wei, Y. Tay, R. Bommasani, C. Raffel, B. Zoph, S. Borgeaud, D. Yogatama, M. Bosma, D. Zhou, D. Metzler, E. H. Chi, T. Hashimoto, O. Vinyals, P. Liang, J. Dean, and W. Fedus,
``Emergent abilities of large language models,'' Transactions on Machine Learning Research (2022).

\bibitem{4} G. M. Edelman, \emph{Neural Darwinism: The Theory of Neuronal Group Selection} (Basic Books, 1989).
\bibitem{5} G. Tononi, O. Sporns, and G. M. Edelman, ``A measure for brain complexity: Relating functional segregation and integration in the nervous system,''
Proc. Natl. Acad. Sci. USA \textbf{91}, 5033-5037 (1994).

\bibitem{6} S. Yao, J. Zhao, D. Yu, N. Du, I. Shafran, K. Narasimhan, and Y. Cao, ``ReAct: Synergizing reasoning and acting in language models,'' arXiv:2210.03629 (2022).
\bibitem{7} T. Webb, K. J. Holyoak, and H. Lu, ``Emergent analogical reasoning in large language models,'' Nat. Hum. Behav. \textbf{7}, 1526-1541 (2023).
\bibitem{8} I. Schlag, T. Irie, and J. Schmidhuber, ``Linear transformers are secretly fast weight programmers,'' arXiv:2102.11174 (2021).
\bibitem{9} A. Katharopoulos, A. Vyas, N. Pappas, and F. Fleuret, ``Transformers are RNNs: Fast autoregressive transformers with linear attention,'' arXiv:2006.16236 (2020).

\bibitem{10} J. Wilting and V. Priesemann, ``25 years of criticality in neuroscience - established results, open controversies, novel concepts,''
Curr. Opin. Neurobiol. \textbf{58}, 105-111 (2019).
\bibitem{11} D. R. Chialvo, ``Emergent complex neural dynamics,'' Nat. Phys. \textbf{6}, 744-750 (2010).
\bibitem{12} J. M. Beggs and D. Plenz, ``Neuronal avalanches in neocortical circuits,'' J. Neurosci. \textbf{23}, 11167-11177 (2003).
\bibitem{13} A. Levina, J. M. Herrmann, and T. Geisel, ``Phase transitions towards criticality in a neural system with adaptive interactions,''
Phys. Rev. Lett. \textbf{102}, 118110 (2009).

\bibitem{14} P. Bak, C. Tang, and K. Wiesenfeld, ``Self-organized criticality: An explanation of $1/f$ noise,'' Phys. Rev. Lett. \textbf{59}, 381-384 (1987).
\bibitem{15} J. T. Stuart, “On the non-linear mechanics of hydrodynamic stability,” J. Fluid Mech. \textbf{4}, 1-21 (1958).
\bibitem{16} J. A. Acebr\'{o}n, L. L. Bonilla, C. J. P\'{e}rez Vicente, F. Ritort, and R. Spigler, ``The Kuramoto model: A simple paradigm for synchronization phenomena,''
Rev. Mod. Phys. \textbf{77}, 137-185 (2005).


\bibitem{17} B. G. Chae, ``A unifield dynamical field theory of learning, inference, and emergence,'' arXiv:2601.10221 (2026).

\bibitem{18} B. G. Chae, ``Recursive dynamics in fast-weights homeostatic reentry networks: Toward reflective intelligence,'' arXiv:2511.06798 (2025).
\bibitem{19} B. G. Chae, ``Continuous-time homeostatic dynamics for reentrant inference models,'' arXiv:2512.05158 (2025).
\bibitem{20} B. G. Chae, ``Renormalization-group geometry of homeostatically regulated reentry networks,'' arXiv:2512.19086 (2025).
\bibitem{21} B. G. Chae, ``Self-organized criticality from protected mean-field dynamics: Loop stability and internal renormalization in reflective neural systems,'' arXiv:2601.04450 (2026).


\bibitem{22} J. J. Hopfield, ``Neural networks and physical systems with emergent collective computational abilities,'' Proc. Natl. Acad. Sci. USA \textbf{79}, 2554-2558 (1982).
\bibitem{23} G. E. Hinton and D. C. Plaut, ``Using fast weights to deblur old memories,'' Proc. 9th Annu. Conf. Cognitive Science Society, 177-186 (1987).
\bibitem{24} J. Schmidhuber, ``Learning to control fast-weight memories: An alternative to dynamic recurrent networks,'' Neural Comput. \textbf{4}, 131-139 (1992).

\bibitem{25} K. Funahashi and Y. Nakamura, ``Approximation of dynamical systems by continuous time recurrent neural networks,''
Neural Networks \textbf{6}, 801-806 (1993).
\bibitem{26} R. Hasani, M. Lechner, A. Amini, D. Rus, and R. Grosu, ``Liquid time-constant networks,'' Proc. AAAI Conf. Artif. Intell. \textbf{35}, 7657-7666 (2021).


\bibitem{27} P. C. Martin, E. D. Siggia, and H. A. Rose, ``Statistical dynamics of classical systems,'' Phys. Rev. A \textbf{8}, 423-437 (1973).
\bibitem{28} H. K. Janssen, ``On a Lagrangian for classical field dynamics and renormalization group calculations of dynamical critical properties,''
Z Phyik B \textbf{23}, 377-380 (1976).
\bibitem{29} C. De Dominicis, ``Techniques de renormalisation de la th\'eorie des champs et dynamique des ph\'enom\`enes critiques,''
J. Phys. Colloq. \textbf{37}, 247-253 (1976).

\bibitem{30} M. M. Churchland, J. P. Cunningham, M. T. Kaufman, J. D. Foster, P. Nuyujukian, S. I. Ryu, and K. V. Shenoy,
``Neural population dynamics during reaching,'' Nature \textbf{487}, 51-56 (2012).
\bibitem{31} J. A. Gallego, M. G. Perich, L. E. Miller, and S. A. Solla, ``Neural manifolds for the control of movement,'' Neuron \textbf{94}, 978-984 (2017).
\bibitem{32} J. P. Cunningham and B. M. Yu, ``Dimensionality reduction for large-scale neural recordings,'' Nat. Neurosci. \textbf{17}, 1500-1509 (2014).

\bibitem{33} D. Sussillo, ``Neural circuits as computational dynamical systems,'' Curr. Opin. Neurobiol. \textbf{25}, 156-163 (2014).
\bibitem{34} V. Mante, D. Sussillo, K. V. Shenoy, and W. T. Newsome, ``Context-dependent computation by recurrent dynamics in prefrontal cortex,'' 
Nature \textbf{503}, 78-84 (2013).

\end{thebibliography}
\end{document}